\pdfoutput=1
\documentclass[aps, rmp, amsmath, a4paper, notitlepage, 11pt, twoside]{revtex4-1}

\usepackage{graphicx}
\usepackage{color}
\usepackage{units}
\usepackage[
font={sf}, 
labelfont={sf,bf},
justification=centerlast,
format=hang,
margin=5mm
]{caption}
\usepackage[
colorlinks,
citecolor=blue,
linkcolor=blue
]{hyperref}

\renewcommand{\i}{\ensuremath{\mathrm i}}

\newcommand{\e}{\ensuremath{\mathrm e}}
\newcommand{\ket}[1]{\left|#1\right>}
\newcommand{\bra}[1]{\left<#1\right|}
\newcommand{\braket}[2]{\ensuremath{\left<#1|#2\right>}}
\newcommand{\TW}[1]{\ensuremath{\unit[#1]{TW/cm^2}}}

\setcitestyle{numbers,square}
\begin{document}

\title{Adiabatic Floquet model for the optical response in femtosecond filaments}

\author{Michael Hofmann}
\email{hofmann@wias-berlin.de}
\affiliation{Weierstrass Institute for Applied Analysis and Stochastics}
\affiliation{Institute for Theoretical Physics at Technical University Berlin}
\author{Carsten Br\'ee}
\affiliation{Weierstrass Institute for Applied Analysis and Stochastics}
\affiliation{Max Born Institute for Nonlinear Optics and Short Pulse Spectroscopy}

\date{\today}

\begin{abstract}
  The standard model of femtosecond filamentation is based on phenomenological assumptions which suggest that the ionization-induced carriers can be treated as free according to the Drude model, while the nonlinear response of the bound carriers follows the all-optical Kerr effect.
  Here, we demonstrate that the additional plasma generated at a multiphoton resonance dominates the saturation of the nonlinear refractive index.
  Since resonances are not captured by the standard model, we propose a modification of the latter in which ionization enhancements can be accounted for by an ionization rate  obtained from non-Hermitian Floquet theory.
  In the adiabatic regime of long pulse envelopes, this augmented standard model is in excellent agreement with direct quantum mechanical simulations.
  Since our proposal maintains the structure of the standard model, it can be easily  incorporated into existing codes of filament simulation.
\end{abstract}

%
%
%
\maketitle
%
%


\section{Introduction}

Femtosecond filaments are a versatile tool of nonlinear optics. 
With the advent of the chirped pulse amplification technique which can deliver ultrashort, intense laser pulses\;\cite{strickland}, filament generation in gaseous media became possible for the first time\;\cite{A.1995}.
Since the optical power in femtosecond pulses can easily exceed the threshold for nonlinear self-focusing,
they can be used to study the catastrophic nonlinear wave collapse\;\cite{Gaeta2000}.
Under loose focusing conditions, however,  the onset of plasma defocusing  may avoid the collapse and lead to the formation of longitudinally extended filaments.
Since filaments have highly directional beam characteristics,  they are theoretically described by the forward Maxwell equation\;\cite{Husakou2001} or the unidirectional pulse propagation equation\;\cite{Kolesik2002}.
Usually, polarization density and current density are plugged into the propagation equation to account for the optical response of the medium.
In the standard model of femtosecond filamentation these quantities are treated phenomenologically\;\cite{L.2007}:
It is assumed that the nonlinear part of the polarization density is governed by the $\chi^{(3)}$-nonlinearity, while the current density due to ionization is described by the Drude model.
Neglecting third harmonic generation, the standard model can be boiled down to a formula for the field-dependent refractive index change according to
\begin{equation}
  \label{eq:stdmod}
  \Delta n(t)= n_2 I(t)-\frac{\rho (t)}{2\rho_c}.
\end{equation}
Here, $n_2$ is the nonlinear refractive index  describing the strength  of Kerr self-focusing, $I(t)$ is the temporal intensity envelope of the laser pulse,
and $\rho$ is the electron density derived from Keldysh theory\;\cite{Keldysh1965}.
Above the critical plasma density $\rho_c$ the electron plasma becomes opaque.
The standard model has proven quite successful in many applications.
However, its validity was questioned by an experimental observation of a saturating Kerr nonlinearity known as higher-order Kerr effect (HOKE)\;\cite{Loriot2009}.
This result triggered  conflicting studies either confirming\;\cite{Bejot2010,Ettoumi2010,Bree2012} or rejecting \;\cite{Polynkin2011,Kosareva2011,Volkova2012} a possible new paradigm of filamentation.

Since there is a growing consensus that the measured deviations from the standard model cannot be described perturbatively via the HOKE model, one may ask if yet another physical mechanism is responsible?
Ab initio simulations of light-matter interaction gave promising explanations which are not accounted for by the standard model.
Among them are trapped population in Rydberg states\;\cite{Volkova2013}, formation of Kramers-Henneberger atoms\;\cite{Richter2013} or Freeman resonances\;\cite{Hofmann2015}.
These findings also shed some light on the conceptual problems of the standard model\;\cite{Kolesik2013}, e.g., the ambiguity in the definition of bound and free electrons in atoms dressed by strong laser fields, or the fact that resonance enhancements are neglected in the ionization rate.
One striking argument is that the perturbative description of the polarization density fails in the regime of filamentation, rendering HOKE questionable\;\cite{Volkova2012,Spott2014}.
Therefore, instead of a phenomenological approach, the demand for an alternative model derived from quantum mechanics is growing.
For instance, one could calculate the polarization density and current density needed to propagate Maxwell's equations by direct numerical simulations of the time-dependent Schr\"odinger equation (TDSE), as done in~\cite{Lorin2012}.
However, this coupled ab initio approach is numerically cumbersome and not feasible for the desired meter-scale propagation distance, the typical longitudinal extension of laboratory-generated filaments.
Therefore, faster computational models for quantum-mechanical calculations of the atomic response have recently been proposed, like the non-local model of~\cite{Rensink2014}, or the approach proposed in~\cite{Kolesik2014} which describes the response in terms of the field-dressed, metastable ground state.

In the first part of this manuscript, we demonstrate the importance of multiphoton resonances which are not accounted for by the standard model.
Our ab initio simulations show that resonances significantly enhance ionization which dominates the nonlinear refractive index.
In the second part, we offer a modification to the standard model with an ionization rate derived from non-Hermitian Floquet theory.
Compared with ab initio calculations, our model accurately reproduces the impact of enhanced ionization in the adiabtic regime of a slowly varying pulse envelope.
The resulting formula for the nonlinear refractive index is evaluated with equal computational costs as the standard model formula of~(\ref{eq:stdmod}) and can be easily integrated into existing numerical routines for filament propagation.


\section{Numerical experiment}

In a recent publication\;\cite{Hofmann2015} the current authors demonstrated that ionization enhancements induced by Freeman resonances\;\cite{Freeman1987a} coincide with a local saturation of the nonlinear refractive index.
This saturation, in turn, was shown to influence  the input power dependence of the clamping intensity and thus the propagation dynamics of optical filaments.
In this section we extend our previous research and prove that resonantly enhanced ionization is indeed responsible for a dramatic decrease of the nonlinear refractive index.
In contrast to our previous work, here we consider the self-induced refractive index modifications seen by a strong pump pulse instead of the cross-induced refractive index change in a pump-probe setup.
The optical response is then compared with a modified version of the standard model, in which the density of free electrons obtained from the direct numerical simulations replaces the electron density calculated according to the Keldysh theory.

\subsection{Time-dependent Schr\"odinger equation}
\label{sec:simulation}

We briefly describe the methods for simulating the optical response of a one-dimensional hydrogen atom under the influence of a strong laser field.
One-dimensional model atoms are popular toy models and give good qualitative predictions of the dynamics of atoms in strong laser fields.
In particular, they were successfully used for theoretical studies of high harmonic generation and atomic stabilization in strong fields\;\cite{hhg,Gavrila2008,Su1996}. 
Starting from first principles, we write down the TDSE in the dipole approximation using atomic units (au),
\begin{equation}
  \label{eq:tdse}
  \i\partial_t\psi(x,t) = \left[-\frac{1}{2}\Delta-\frac{1}{\sqrt{x^2+\alpha^2}}+E(t)x \right] {\psi(x,t)}.
\end{equation}
To circumvent the singularity of the atomic potential at the origin, a soft-core potential is used which yields the correct ionization potential of hydrogen, $I_p=\unit[13.6]{eV}\equiv\unit[0.5]{au}$ if  $\alpha=\sqrt{2}$.
For integrating the TDSE numerically we discretize time $t\rightarrow t_n$ and space $x\rightarrow x_j$ with equal spacings $\tau=\Delta t=\unit[0.01]{au}$ and $h=\Delta x=\unit[0.1]{au}$, respectively.
For the second derivative in the Hamiltonian we can write the central difference
\begin{equation}
  \label{eq:laplace}
  \Delta\psi \approx \frac{\psi_{j-1}-2\psi_j+\psi_{j+1}}{h^2},
\end{equation}
where $\psi^n_j$ is the wave function at time step $t_n$ on the grid point $x_j$.
Thus, the matrix representing $H$ on the grid has a tridiagonal structure.
The implicit Crank-Nicolson propagator is commonly chosen to compute the wave function for the next time step,
\begin{equation}
  \label{eq:crank-nicolson}
  U(t_{n+1},t_n) = \left( 1 + \i\frac{\tau}{2} H(t_{n+1}) \right)^{-1} \left( 1 - \i\frac{\tau}{2} H(t_n) \right),
\end{equation}
which preserves unitarity and is second order accurate in space and time\;\cite{Goldberg1967}.
When the propagator is applied to the wave function, i.e., $\psi^{n+1} = U(t_{n+1},t_n)\psi^n$, we obtain a set of linear equations,
\begin{equation}
  \label{eq:linear-eq}
  M^+_{n+1}\psi^{n+1} = M_n^-\psi^n \quad \text{with} \quad M^\pm_n \equiv \left( 1 \pm \i\frac{\tau}{2} H(t_{n}) \right).
\end{equation}
These equations can be efficiently solved using the tridiagonal matrix algorithm\;\cite{NumRecipes1993}.
As an initial condition we provide that $\psi^0$ is the atom's ground state.
Reflections at the spatial ends of the integration domain are suppressed by a \unit[40]{au} wide absorbing boundary layer\;\cite{Manolopoulos2002}.

\subsection{Nonlinear refractive index and resonantly enhanced ionization}
\label{sec:obs}

 We consider a laser pulse with a central wavelength of \unit[800]{nm}, which corresponds to an angular frequency of $\omega_0=\unit[0.057]{au}$ and an optical cycle length of $T=\unit[2.67]{fs}\equiv\unit[110.3]{au}$. 
To ensure that the electric field $E(t)$ has vanishing dc component, we introduce the vector potential $A(t)$, from which we reconstruct the electric field according to $E(t)=-\partial_t A(t)$.
The vector potential has a cosine square envelope,
\begin{equation}
  \label{eq:vector-field}
  A(t)=  A_0\cos^2(\pi t/T_p)\cos(\omega_0 t) \quad \text{for} \quad |t| \le T_p/2, 
\end{equation}
where $T_p$ is the total pulse duration, and the peak amplitudes of $E$ and $A$ are related via $E_0=\omega_0 A_0$.
We already know from our previous work that Freeman resonances between laser-dressed states lead to drops in the refractive index, and that those resonances are more pronounced when a long flat-top pulse is used.
We therefore additionally employ a flat-top pulse with a four-cycle ramp up and down, respectively, according to~(\ref{eq:vector-field}) but with an 80-cycle long central region of constant amplitude in between.

Having solved the TDSE for a single electron, the polarization density for a gas of hydrogen atoms at standard condition, with number density $\sigma=\unit[4\cdot10^{-6}]{au}$, can be obtained from the atomic dipole moment,
\begin{equation}
  \label{eq:polarization}
  P(t_n)=-\sigma \bra{\psi^n}x\ket{\psi^n}.
\end{equation}
Denoting  $\hat X(\omega)$ as the Fourier transform of $X(t)$, we can define a time-averaged susceptibility and thus a refractive index of the hydrogen gas interacting with the laser pulse,
\begin{equation}
  \label{eq:susceptibility}
  \chi = \frac{\hat P(\omega_0)}{\epsilon_0\hat E(\omega_0)}, \quad  n = \mathrm{Re}\left(\sqrt{1+\chi}\;\right).
\end{equation}
By repeating the simulation with variable peak intensity $I$ up to \TW{40}, we obtain an intensity-dependent, nonlinear refractive index
\begin{equation}
  \label{eq:refindex}
  \Delta n(I) = n(I) - \lim\limits_{I\rightarrow0}n(I).
\end{equation}
To ensure $n(I)$ correctly captures the response from ionized electrons, the simulation volume is constantly adjusted with increasing $I$.
Less than 0.1\,\% of ionization is absorbed at the boundaries.

\begin{figure}[tb]
  \centering
  \includegraphics[]{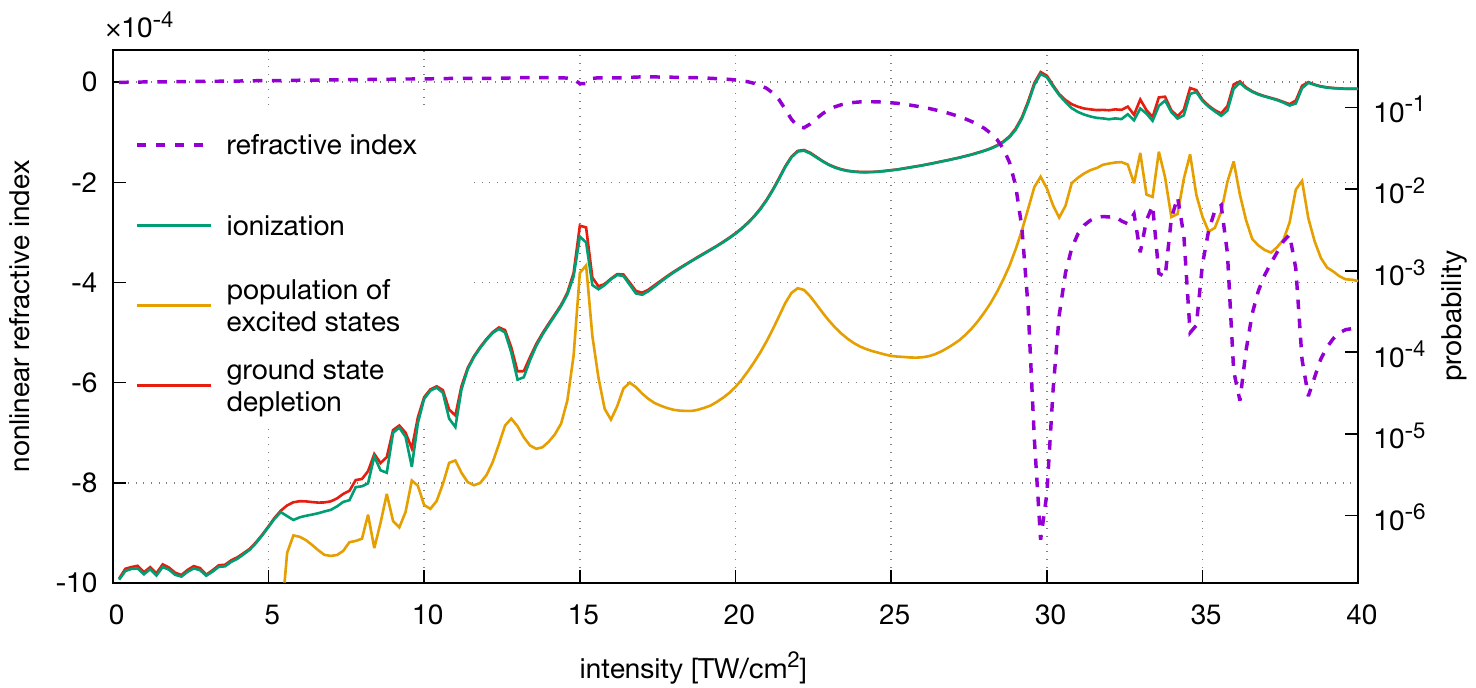}
  \caption{TDSE results for a 80-cycle flat-top pulse. Left axis: nonlinear refractive index (purple, dashed). Right axis: Probability of ionization (green), excited states' population (orange) and ground state depletion (red) at the end of the pulse.}
  \label{fig:nri_vs_rho}
\end{figure}

Our next objective is to seek for correlations between the nonlinear refractive index and the probability of ionization, the population of excited states and the ground state depletion.
Thus, we compute the probability that the electron occupies a specific field-free bound state after the pulse is gone.
This can be extracted directly from the wave function at the end of the pulse by projecting onto those eigenvectors of the discrete Hamiltonian which correspond to bound atomic states, denoted as $\phi_k$,
\begin{equation}
  \label{eq:boundstates}
  \beta_k = |\braket{\phi_k}{\psi(T_p/2)}|^2 \quad\text{and}\quad \beta_\text{bound} = \sum_{k} \beta_k.
\end{equation}
In practice, the sum in~(\ref{eq:boundstates}) includes the first 100 bound states which is enough to ensure convergence of $\beta_\text{bound}$.
Obviously, the ground state depletion is $1-\beta_0$, and the probability for the electron to be in an excited state is $\beta_\text{bound}-\beta_0$.
The ionization probability is defined as the non-bound part of the wave function, $\beta_\text{free}= 1 - \beta_\text{bound}$.

Results for the 80-cycle flat-top pulse with $T_\text{FWHM}=\unit[215.8]{fs}$ are shown in figure~\ref{fig:nri_vs_rho}.
The nonlinear refractive index (purple) exhibits local drops around peak intensities of 15, 22 and \TW{30} as well as above \TW{31}.
Apparently, these drops coincide with an increased ground state depletion (red).
The depletion is mainly caused by enhanced transitions into the continuum (green), since the probability to end up in an excited bound state (orange) is 1--2 orders of magnitude smaller.
The origin of the enhanced ionization probability are Freeman resonances identified in~\cite{Hofmann2015}.
To consolidate the existence of those resonances, we briefly sketch how they occur:
Since the ground state of an atom is strongly bound, its energy remains almost constant when an external field is applied.
With increasing intensity, however, the ac-Stark effect shifts the energy levels of the dressed Rydberg states by the ponderomotive potential, $U_p=E^2/(4\omega_0^2)$.
Eventually, a Freeman resonance arises when the energy difference between a Rydberg state and the ground state is a multiple of the photon energy.
These resonances are usually preceded by a channel closure, which occurs when the ground state is in multiphoton resonance with the ponderomotively up-shifted continuum threshold.
In consequence, we expect an enhanced population of the participating Rydberg state at a Freeman resonance.
Indeed, above the 9-photon channel closure at \TW{5.5} in figure~\ref{fig:nri_vs_rho}, we observe an increase and successive peaks in the occupation probability of the excited states.
Interestingly, this seems to be anticorrelated to the ionization probability, which is decreased at the same intensities, most likely due to interference stabilization\;\cite{Popov2003}.
This stabilization vanishes for peak intensities $\ge\TW{15}$, where transitions into both bound and continuum states are resonantly enhanced.

\begin{figure}[tb]
  \centering
  \includegraphics[]{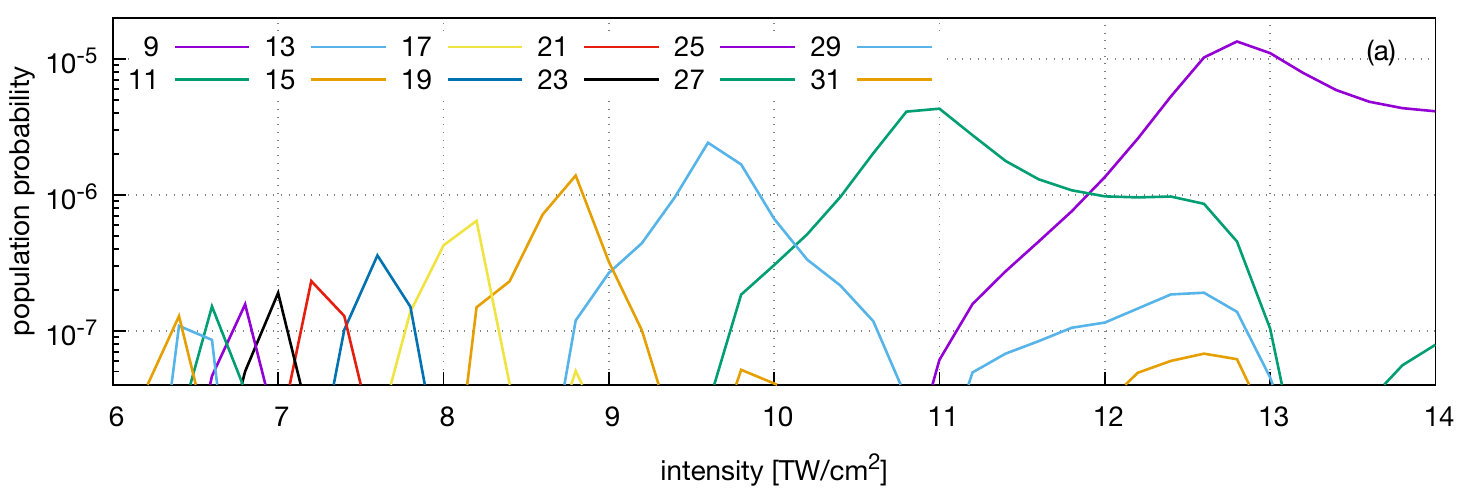} \\[2mm]
  \includegraphics[]{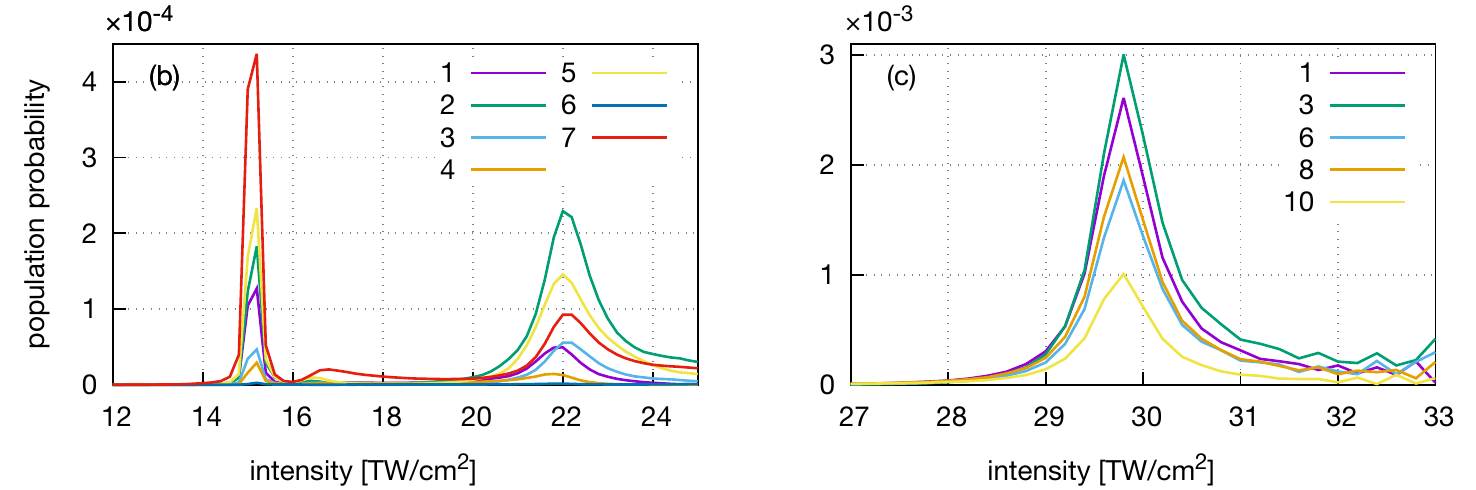} \\[2mm]
  \includegraphics[]{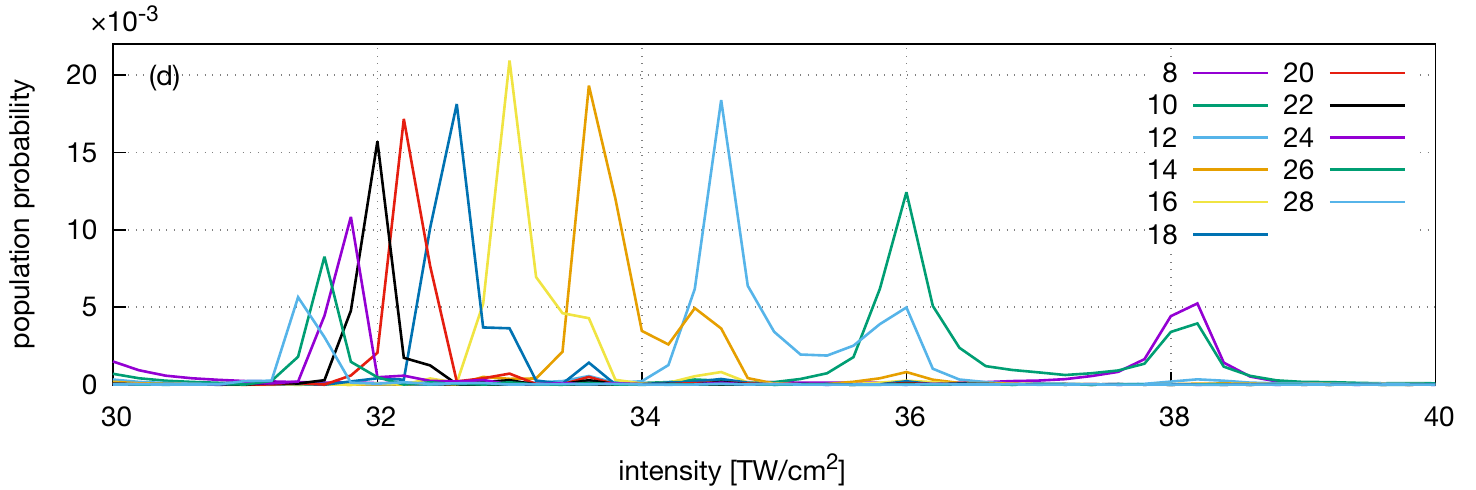}
  \caption{
    Population transfer to excited states (given by the number in the legend) after the pulse is gone. 
    Only significantly populated states are shown.
    Enhancements give evidence of resonances between field-dressed states: 
    (a) and (d) Freeman resonances after the 9- and 10-photon channel closure, respectively.
    (b) and (c) ``Mixed resonances'' involving multiple excited states. 
  }
  \label{fig:population}
\end{figure}

We now look closer at individual Rydberg states which should be dominantly populated at a Freeman resonance.
As shown in figure~\ref{fig:population}(a), we resolve transitions up to the 31st excited state just after the 9-photon channel closure.
Note that parity conservation prohibits a 9-photon transition from the even ground state to even excited states, thus only transitions to odd-parity states are allowed.
After the 10-photon channel closure at \TW{30.2} in figure~\ref{fig:population}(d), we again observe Freeman resonances, now with 10-photon transitions to even-parity excited states.
Somehow different are the resonances near 15, 22 and \TW{30} in figure~\ref{fig:population}(b) and (c):
Increasing the peak intensity beyond the range of figure~\ref{fig:population}(a), we expect transitions to the 7th, 5th and 3rd excited state, since with increasing intensity lower lying excited states come into multiphoton resonance with the ground state.
However, these transitions are not dominant since the population of other excited states is of same magnitude. 
Obviously, multiple excited states  contribute to these resonances.
This may be related to the fact that the energy shift of lower lying excited states shows a more complicated behavior than that of Rydberg states.
Resulting interactions between dressed excited states may lead to a parity change, which explains that also even-numbered excited states are occupied.
Nevertheless, these ``mixed resonances'' facilitate enhanced ionization as seen in figure~\ref{fig:nri_vs_rho}.

\subsection{Modification of the standard model}
\label{sec:mod_std}

The observations of the last paragraph clearly indicated that resonances in the refractive index are correlated to the ground state depletion.
Our results also revealed that the ground state depletion is dominated by transitions into the continuum.
They therefore confirm the standard model assumption that the refractive index saturation stems from free electrons.
A significant difference to the standard model, however, lies in the physical mechanism underlying the ionization enhancement which we identified as resonances between laser-dressed states.
Thus, the standard model should be able to reproduce our simulation results if the resonantly enhanced ionization is taken into account.
To test this assumption, we write down a modified, time-independent version of the nonlinear refractive index~(\ref{eq:stdmod}),
\begin{equation}
  \label{eq:mod_std_model}
  \Delta n(I) = n_2I - \frac{\tilde\rho(I)}{2\rho_c},
\end{equation}
where $I$ is the peak intensity of the laser pulse. Furthermore, we have to take into account that the density of free carriers is unambiguously defined only for vanishing external electric field\;\cite{Bejot2013}.
We therefore replaced in~(\ref{eq:stdmod}) the Keldysh-like ionization density $\rho(t)$ with the electron density at the end of the pulse $\tilde\rho(I)$, calculated from the direct numerical simulations.
We then compare three different cases:
\begin{equation}
  \label{eq:rho}
  \tilde\rho(I) = \eta\,\sigma
    \begin{cases}
      \beta_\text{free}(I) & \text{case A, ionization only,} \\
      \beta_\text{free}(I) + \beta_\text{Ryd}(I) & \text{case B, ionization and Rydberg states,} \\
      1-\beta_0(I) & \text{case C, full ground state depletion.}
    \end{cases}
\end{equation}
Since this model disregards the temporal evolution of the plasma density and uses the density at the end of the pulse instead, we introduce a phenomenological correction factor $\eta\approx0.5$ which depends on the field envelope.
Case~B is motivated by the fact that the contribution to the refractive index from populated Rydberg states is similar to that of free electrons\;\cite{Fedorov1997}.
Additionally, electrons which are freed during the pulse may be captured back into Rydberg states when the field is gone.
Here, we consider all states above the ninth excited state as Rydberg states since their corresponding Floquet energies shift with the ponderomotive potential (cf. figure~1 in~\cite{Hofmann2015}). 
The probability to find the atom in a Rydberg state is then given by $\beta_{\text{Ryd}}=\sum_{k>9} \beta_k$.
The overall population of excited states, however, is small compared to the ionization probability as seen in figure~\ref{fig:nri_vs_rho}.
Therefore, case~C  may serve as a simple, but accurate approximation. 

\begin{figure}[t]
  \centering
  \includegraphics[]{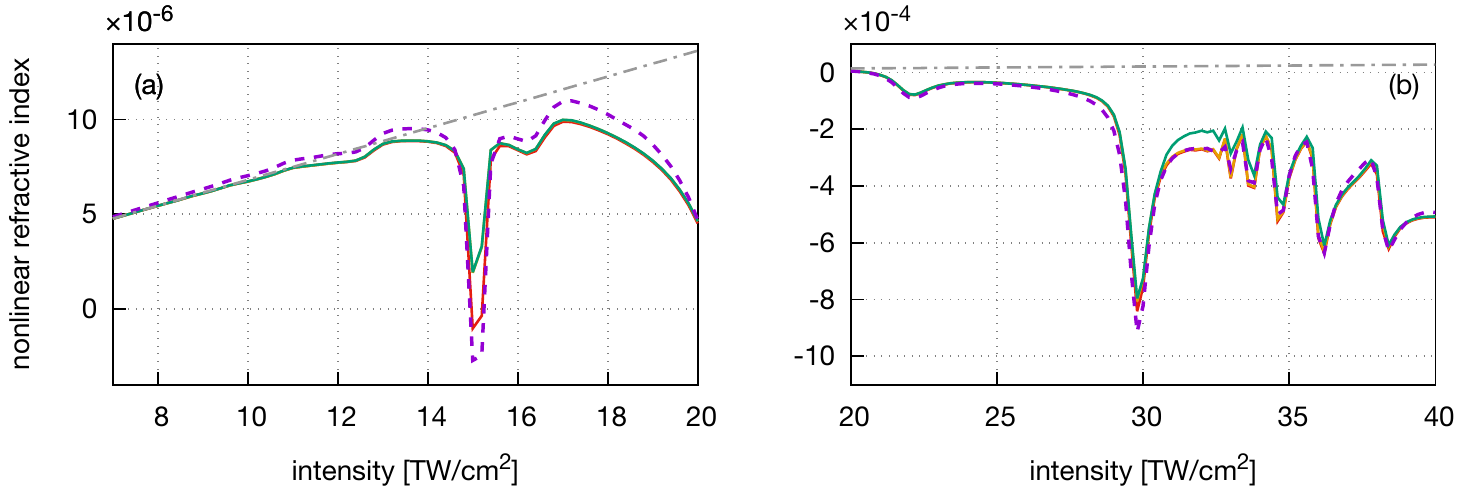} \\
  \includegraphics[]{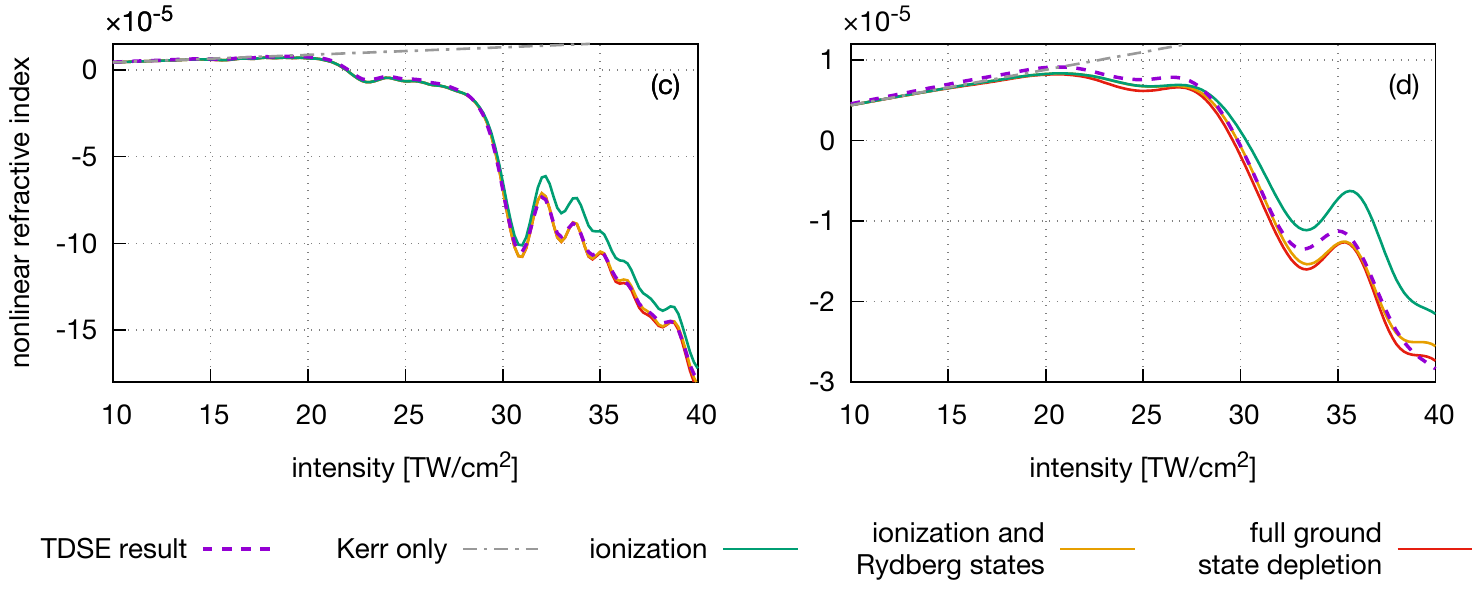}
  \caption{Comparison of the nonlinear refractive index from TDSE (purple, dashed) with the modified standard model~(\ref{eq:mod_std_model}), cases A (green), B (orange) and C (red). The dash-dotted line indicates the extrapolated, linear Kerr contribution. (a) and (b) 80-cycle flat-top pulse with $\eta=0.41$. (c) and (d) $\cos^2$-pulse with 96 and 24 cycles, $\eta=0.5$ and 0.49, respectively. }
  \label{fig:mod_vs_std}
\end{figure}

Results are presented in figure~\ref{fig:mod_vs_std}, together with the exact TDSE calculation~(\ref{eq:refindex}), for different pulse shapes and lengths.
First, we employed the 80-cycle flat-top pulse.
For low peak intensities shown in panel~(a), the refractive index for both model (A: green, B: orange, C: red) and simulation (purple, dashed) follow the Kerr-nonlinearity $n_2I$ (grey, dash-dotted) until enhanced ionization becomes non-negligible.
As an interesting side note, the TDSE result is obviously superlinear, indicating that the next coefficient in a perturbative description, $n_4$, has to be positive.
Although there is a small deviation above \TW{10} between model and simulation, the agreement between the two is astounding close.
Only for the \TW{15} resonance, the drop in the refractive index is significantly larger than predicted by case A and B (which overlap in this panel).
Case C, on the other hand, is closer to the simulation result.
This observation also holds for resonances at \TW{22} and \TW{29.8} in panel~(b) which shows the full intensity range on a larger scale for the same flat-top pulse.
All in all the deviation remains small.
After the 10-photon channel closure, the increased population of Rydberg states (cf. figure~\ref{fig:population}) becomes noticeable in the refractive index.
As expected, case A (ionization only) diverges from the simulation while case B and C are in good agreement.
Model and simulation astonishingly coincide even for a purely $\cos^2$-pulse as shown in panel~(c) and (d), where the total pulse duration is 96 cycles ($T_\text{FWHM}=\unit[93.3]{fs}$) and 24 cycles ($T_\text{FWHM}=\unit[23.3]{fs}$), respectively.
Again, we notice deviations in case A for peak intensities larger than \TW{30}, but case B and C are close to the numerical result.
Overall, our modification of the standard model is in remarkable agreement with the nonlinear refractive index from pure TDSE calculations, manifesting the conclusion that resonantly enhanced ionization is the driving component in the saturation and sign inversion.
%


\section{Adiabatic Floquet resonance model for the nonlinear optical response}

In this part of the manuscript, the results of our numerical studies will be cast into an augmented standard model of the optical response in filaments which can be easily built into existing numerical simulation codes for femtosecond filamentation.
Our proposed adiabatic Floquet resonance (AFR) model is based on the following observations: 
First, we have convinced ourselves and hopefully the reader that the optical response is mainly determined by resonantly enhanced transitions from the ground state and the resulting ground state depletion. 
Second, we have shown that the Kerr part of the standard model is nearly unaffected by the considered non-perturbative effects. 
In consequence, our model for the time-dependent nonlinear refractive index is
\begin{equation}
  \label{eq:aug_model}
  \Delta n^\mathsf{AFR}(t)=n_2I(t)-\frac{\rho_0(t)}{2\rho_c},
\end{equation}
where $\rho_0(t)=\sigma(1-\beta_0(t))$ is the macroscopic ground state depletion.
It is governed by a rate equation
\begin{equation}
  \label{eq:rate_eq}
  \partial_t\rho_0(t)=w[I(t)](\sigma-\rho_0(t)),
\end{equation}
with $w[I]$ as the intensity-dependent transition rate to excited and continuum states.
In principle, $\rho_0(t)$ has to be extracted from direct numerical simulations of the TDSE, and nothing would be gained. 
Nevertheless, in the adiabatic regime of a slowly varying pulse envelope, we can make use of non-Hermitian Floquet theory to calculate the transition rate $w[I]$. 
Since in the adiabatic case, no transitions to excited states are observed, $w[I]$ is here a pure ionization rate and $\rho_0$ a free electron density, in complete correspondence to the usual standard model. 
For our AFR model it is therefore sufficient to replace the Keldysh-type ionization rates in the standard model by an ionization rate obtained from the non-Hermitian Floquet theory.

\subsection{Calculation of Floquet multipliers and eigenstates}

Non-Hermitian Floquet theory describes the atom dynamics in terms of Floquet resonances\;\cite{Moiseyev1998}. 
These are metastable electronic states in a finite volume which are subject to outgoing wave boundary conditions and thus take ionization effects into account.
In this approach, the ionization rate is identical to the inverse lifetime of the Floquet ground state resonance.
To ensure $L^2$-integrability of the resonance wave functions, we employed complex rotation $x\rightarrow x \e^{\i\theta}$ to the Hamiltonian
\begin{equation}
  \label{eq:H_NH}
 H(x) = -\frac{1}{2}\Delta-\frac{1}{\sqrt{x^2+\alpha^2}}-iA(t)\frac{\partial}{\partial x}.
\end{equation}
Here, we opted for the velocity gauge, since in the length gauge, complex rotation  would result in a blow-up of the wave function during time propagation.
Also, we boosted the numerical accuracy of the discretized spatial derivatives to fourth order by employing the Numerov approximation employed in~\cite{muller1999}.
By increasing the order of our numerical scheme, we ensure an accurate determination of ionization rates and Floquet energies.
The ionization rate and Floquet resonances can be obtained from the Floquet eigenvalue equation
\begin{equation}
 (H(x\e^{\i\theta})-i\partial_t)\phi(x,t)=\epsilon\phi(x,t),
 \label{eq:floquet}
\end{equation}
where $\epsilon$ is the complex quasi-energy and $\Gamma=-2\,\mathrm{Im}\,\epsilon$ is the inverse lifetime of a resonance. 
The Floquet wave function $\Psi=\e^{-\i\epsilon t}\phi$ is quasi-periodic with
\begin{equation}
 \Psi(x,t+T)=\e^{-\i\epsilon T}\Psi(x,t),
 \label{eq:quasiperiodic}
\end{equation}
where $T=2\pi/\omega_0$ is period of the laser pulse.
$\Psi$ decays exponentially due to the negative imaginary part of the quasi-energy $\epsilon$ ($\Gamma>0$).
Typically, the ground state resonance $\phi_0$ is the most stable eigensolution with the smallest decay rate $\Gamma_0$, which can be interpreted as the rate of ionization from the ground state. 
This is exactly the quantity we are interested in to build our augmented standard model.
Instead of solving~(\ref{eq:floquet}) directly, we note that by substituting $\Psi(x,t+T)= U(T)\,\Psi(x,t)$ into (\ref{eq:quasiperiodic}), it can be rewritten as an eigenvalue equation for the one-cycle propagator $\mathcal U \equiv U(T)$.
Consequently, the Floquet wave functions are eigenfunctions of ${\mathcal U}$, whereas the Floquet multipliers $\e^{-\i\epsilon T}$ are the corresponding eigenvalues. 
For diagonalization of $\mathcal U$ we employ the eigenvectors  $\chi_n$ of the discrete counterpart of the free particle Hamiltonian $H_0=-\Delta/2$. By propagating $\chi_n$ along one optical cycle we calculate the matrix elements
\begin{equation}
  \mathcal{U}_{mn}=\bra{\chi_m(0)}\mathcal U\ket{\chi_n(0)} = \langle\chi_m(0)|\chi_n(T)\rangle.
\end{equation}
Diagonalization of $\mathcal U$ then yields the complex Floquet multipliers $e^{-i\epsilon_{\alpha}T}$ and the corresponding Floquet states $\phi_{\alpha}$. 
Note that in the analytic theory, according to the Balslev-Combes theorem\;\cite{ABCtheorem1,ABCtheorem2}, the quasi-energies of the quasi-bound resonances do not depend on the chosen rotation angle $\theta$. 
However, this does not necessarily hold true for the numerical treatment, since one usually diagonalizes $\mathcal U$ using either a truncated set of basis functions or a finite space grid. 
In these cases, the rotation angle has to be chosen such that the quasi-energies become stationary with respect to a variation of $\theta$. 
In our case, we obtain reasonable results for $\theta=0.3$.

\begin{figure}[tb]
  \centering
  \includegraphics[]{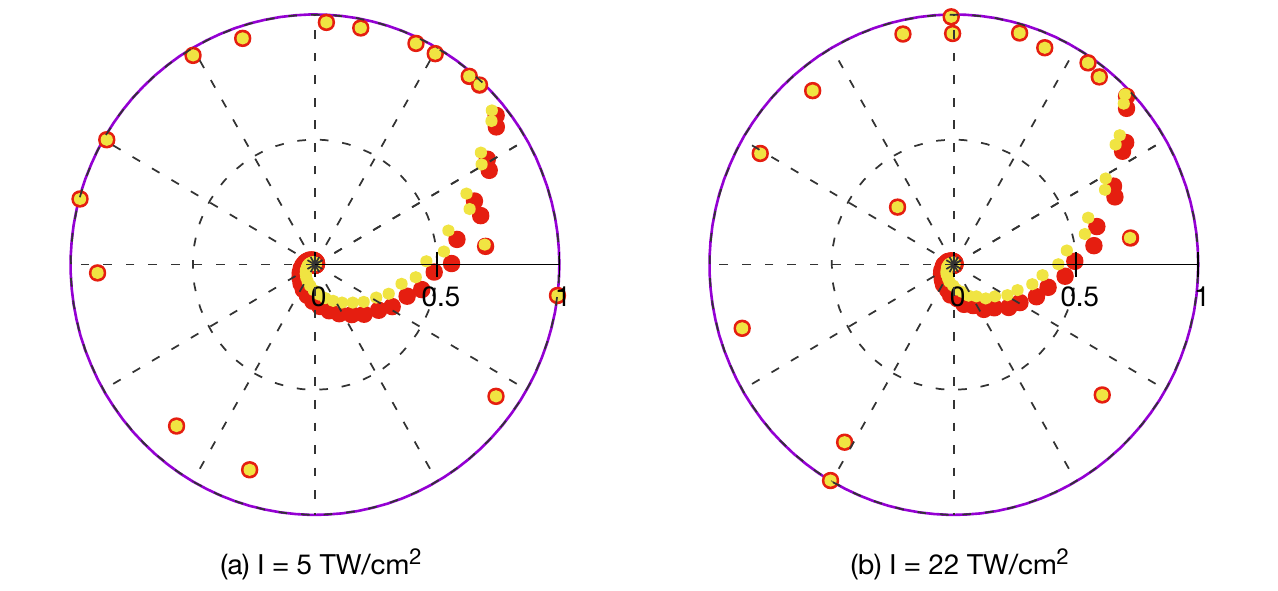}
  \caption{
    Polar plot of the Floquet multipliers $\e^{-\i\epsilon T}$ at two different intensities.
    Red dots: complex rotation $x\,\e^{\i\theta}$ with angle $\theta=0.30$. 
    Yellow dots: $\theta=0.34$.
    Floquet resonances are stable w.r.t.~$\theta$-variation and can therefore be identified as superimposed red an yellow dots.
    A full rotation corresponds to an energy difference of one photon energy.
    The distance to the unit circle (purple) is a measure for the decay rate of a resonance.
  }
  \label{fig:floquet_spiral}
\end{figure}

In figure~\ref{fig:floquet_spiral}, we plot the Floquet multipliers $\e^{-\i\epsilon T}$ in the complex plane for two different values of $\theta$. 
The multipliers corresponding to the discrete quasi-continuum can be observed to spiral into the origin. 
In accordance with the Balslev-Combes theorem, the continuum energies are transformed as $E_c\rightarrow E_c \,\e^{-2\i\theta}$, and the positions of the corresponding Floquet multipliers consequently depend on the rotation angle $\theta$.
In contrast, the positions of the  multipliers corresponding to Floquet resonances are stationary w.r.t. $\theta$-variation.
Therefore, a small change in $\theta$ allows for a straightforward identification of the Floquet resonances.
Furthermore, the distance of the Floquet multipliers to the unit circle is a measure of the lifetime $\Gamma$. 
While the Floquet multipliers of stable bound states ($\Gamma=0$) are located on the contour $\mathcal{C}$ of the unit circle, the lifetime $\tau=1/\Gamma$ of a resonance decreases with increasing distance from $\mathcal{C}$. 
Going from $I=\TW{5}$ in figure~\ref{fig:floquet_spiral}(a) to \TW{22} in (b), resonances move closer to the origin for higher intensities, i.e., they are less stable and ionization is increased.
Additionally, in panel (b), we observe resonances with almost the same complex angle in the polar plot.
Since a full rotation corresponds to an energy difference of one photon energy, these resonances facilitate multiphoton transitions.

To calculate an intrapulse ionization rate $\Gamma_0$ for the ground state resonance, we solve the eigenvalue equation~(\ref{eq:quasiperiodic}) for a set of intensities up to \TW{40}. 
When the intensity increases above a channel closure, $\Gamma_0$ exhibits pronounced peaks stemming from multiphoton or Freeman resonances, as shown in figure~\ref{fig:ion_vs_tdse}(a).
In principle, the obtained ionization rate is valid only in the adiabatic regime of a slowly varying pulse envelope.
In this case, resonance enhanced ionization transfers an electron in the ground state resonance directly into the continuum, without populating the intermediate resonant state, as pointed out in~\cite{Freeman1992}. 
However, as the pulse approaches the few-cycle regime, population transfer to excited Floquet resonances has to be taken into account, and the pronounced resonance peaks in $\Gamma_0(I)$ are expected to smear out for shorter pulses.

\subsection{Augmented standard model with ionization rate from non-Hermitian Floquet theory}

Having calculated the ionization rate $\Gamma_0$, we can compare the predictions from our AFR model~(\ref{eq:aug_model}) and~(\ref{eq:rate_eq}) with the results of direct numerical simulations.  
%
First, we solve the TDSE to obtain the polarization density $P(t)$ for the given electric field $E(t)$. 
From these quantities, a time dependent susceptibility $\chi(t)$ is computed by forming the ratio of the corresponding complex analytic signals\;\cite{Brown2012},
\begin{equation}
 \chi(t)=\frac{\mathcal{P}(t)}{\epsilon_0\mathcal{E}(t)}
 \label{eq:dn_num}
\end{equation}
with $\mathcal{P} = P + \i\mathcal H(P)$ (and likewise $\mathcal E$), and $\mathcal H$ is the Hilbert transform.
From~(\ref{eq:dn_num}), the (ab initio) nonlinear refractive index is obtained as $\Delta n(t)=\sqrt{1+\chi(t)}-n_0$, where $n_0$ is the field-free background refractive index.
In order to obtain the corresponding predictions of the AFR model, denoted as $\Delta n^\mathsf{AFR}(t)$, we have to determine the nonlinear index $n_2$:
For intensities below \TW{1}, i.e., in the perturbative Kerr regime, the refractive index reproduces the intensity profile of the pulse, $\Delta n(t)=n_2 I(t)$.
After evaluating $\Delta n(t)$ from TDSE at the local maxima of $I(t)$ for a series of low peak intensities, a linear fit yields $n_2 = 6.5\times 10^{-7}$\,cm$^2/$TW.
This allows the evaluation of the Kerr part of the nonlinear refractive index $\Delta n^\mathsf{AFR}(t)$.
 The plasma part in~(\ref{eq:aug_model}) can be calculated by plugging the ionization rate $\Gamma_0$ into the rate equation~(\ref{eq:rate_eq}) and by integrating the latter for temporal intensity profiles $I(t)$.

\begin{figure}[tb]
  \centering
  \includegraphics[]{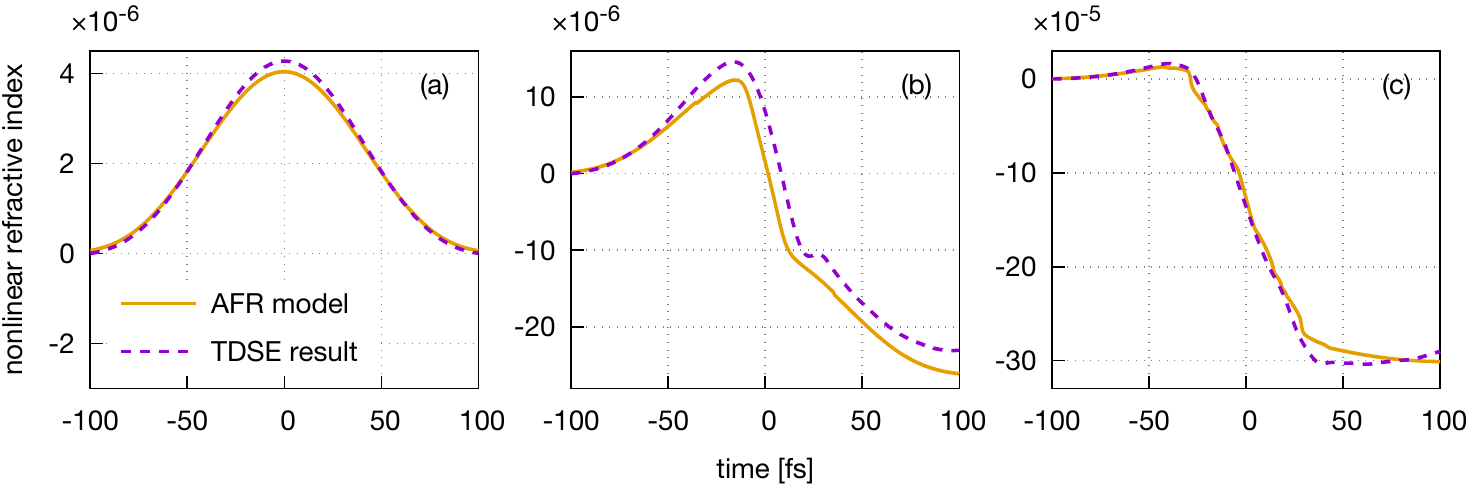}
  \caption{Time-dependent nonlinear refractive index $\Delta n(t)$ for a 96-cycle $\cos^2$-pulse with peak intensities $I=\TW{6.2}$~(a), \TW{22.2}~(b) and \TW{38.2}~(c). Solid lines: $\Delta n^\mathsf{AFR}(t)$ according to the AFR model, (\ref{eq:aug_model}) and~(\ref{eq:rate_eq}).
  Dashed lines: $\Delta n(t)$ from direct numerical simulations using~(\ref{eq:dn_num}).}
  \label{fig:afr_vs_tdse}
\end{figure}

Figure~\ref{fig:afr_vs_tdse} shows temporally resolved refractive indices for peak intensities 6.2, 22.2 and \TW{38.2} using a $\cos^2$-pulse with 96 cycles.
If we compare the AFR results (orange) to that of the direct simulations (purple, dashed), we find that our model gives excellent predictions for the nonlinear refractive index:
We clearly see that in the leading edge the pulse profile is recovered in accordance with the Kerr nonlinearity $\Delta n(t)=n_2I(t)$, panel (a) and (b).
However, as plasma accumulates during the laser-atom interaction, the refractive index correction in the trailing edge of the pulse becomes more negative and turns the nonlinearity into a defocusing one, panel (b) and (c).
This behavior is also well-known from the standard model and responsible for the dynamic spatial replenishment scenario described in~\cite{replenishment}.

\begin{figure}[tb]
  \centering
  \includegraphics[]{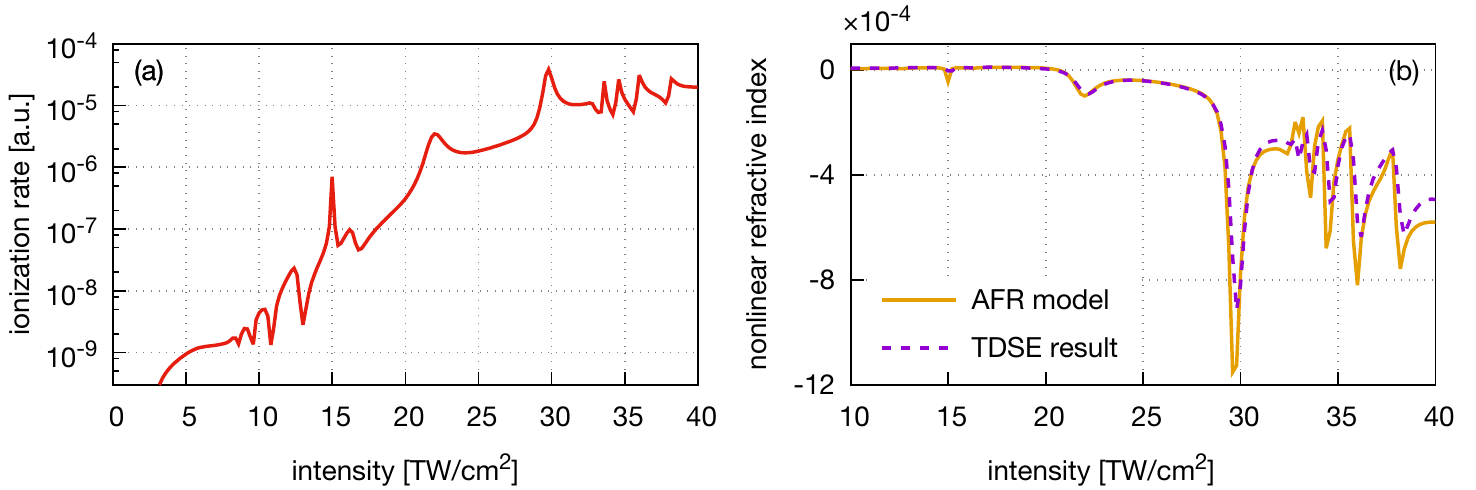} \\
  \includegraphics[]{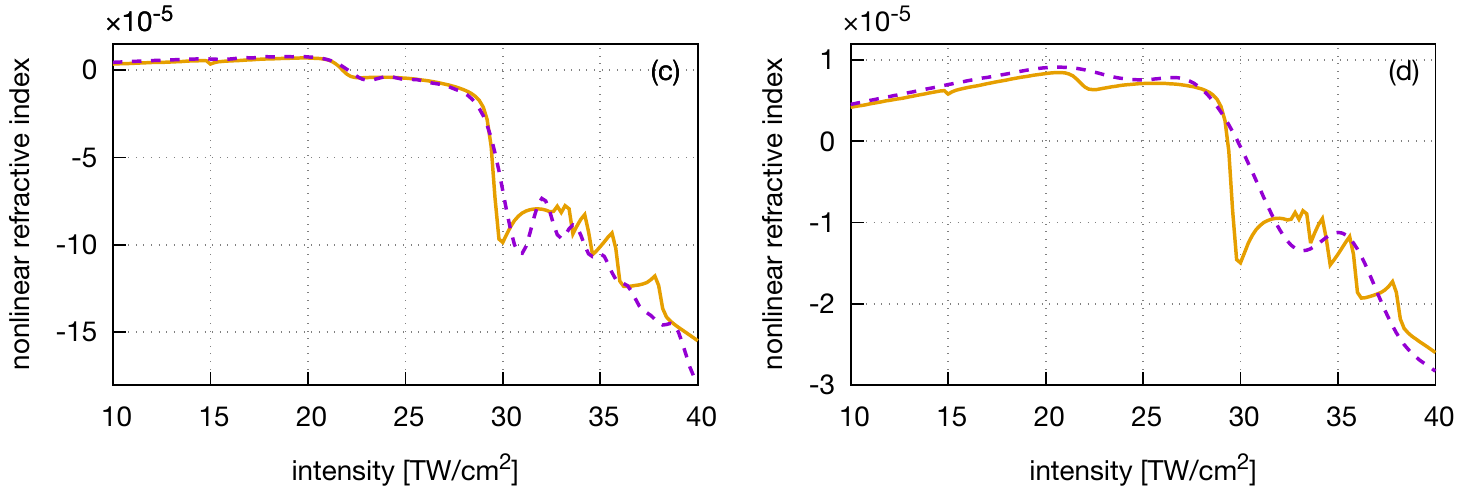}
  \caption{
    Ionization rate $\Gamma_0$ of the ground state Floquet resonance versus intensity~(a). 
    Intensity-dependent refractive index for a 80-cycle flat-top pulse~(b) or  $\cos^2$-pulses with 96~(c) and 24 optical cycles~(d).}
  \label{fig:ion_vs_tdse}
\end{figure}

To conclude our comparison with TDSE results, we derive the intensity dependence of the refractive index as predicted by the AFR model.
In particular, if we know the polarization density $P^\mathsf{AFR}(t)$, we can employ~(\ref{eq:susceptibility}) to calculate the time-averaged susceptibility $\chi^\mathsf{AFR}(\omega_0)$ at the center frequency of the pulse.
In order to construct $P^\mathsf{AFR}(t)$, we use the general relation
\begin{equation}
  P(t)=\epsilon_0\chi(t)E(t) \quad\text{and}\quad \chi(t)=n(t)^2-1,
\end{equation}
where, in this case, $n(t)=n_0+\Delta n^\mathsf{AFR}(t)$.
The nonlinear refractive index $\Delta n^\mathsf{AFR}(I)$ then directly follows from~(\ref{eq:susceptibility}) upon Fourier transform of $P^\mathsf{AFR}(t)$ and $E(t)$  with various peak intensities~$I$.

In figure~\ref{fig:ion_vs_tdse}(b)--(d), we show the intensity-dependent refractive indices obtained from both model and simulation versus peak intensity~$I$: for a flat-top comprising of 80 optical cycles~(b) as well as for $\cos^2$-envelopes with $T_p=96\,T$~(c) and $T_p=24\,T$~(d).
For the long flat-top pulse, the agreement between the direct numerical calculation and the augmented standard model are indeed excellent. 
In fact, for long pulses we can employ the adiabatic approximation with respect to the pulse envelope:
Here, the wave function remains in the ground state resonance which adiabatically adapts to the instantaneous intensity.
In this case, the ionization rate $\Gamma_0$ accurately describes the actual ionization from the ground state resonance.
Moving to shorter pulses in figures~\ref{fig:ion_vs_tdse}(c) and (d), however, we see deviations  between the two calculations. 
Especially, above \TW{30}, the AFR refractive index exhibits strong resonances, which appear smeared out in the $\Delta n(I)$ from the direct simulations. 
Obviously, for shorter pulses, the evolution of the envelope can no longer be considered adiabatic.
In consequence, transitions occur from the ground state resonance to higher lying resonances, and the ionization rate $\Gamma_0$ becomes less accurate.


\section{Conclusion}

In our manuscript, we propose an augmented standard model for the phenomenological description of the optical response in femtosecond filaments. 
Our AFR model maintains the structure of the standard model, which is written as a sum of the Kerr-induced refractive index change and a plasma-induced one, where the time-dependence of the plasma generation is governed by an ionization rate $w[I]$. 
However, instead of an ionization rate derived from Keldysh theory, we employ an ionization rate obtained from non-Hermitian Floquet theory. 

The development of our model was led by numerical experiments which clearly showed that resonantly enhanced ionization has a strong impact on the nonlinear refractive index modifications. 
Interestingly, drops in the nonlinear susceptibility or polarizability have already been noted  in~\cite{Nurhuda2002,Kohler2013,Richter2013}.
Although, in the above-mentioned works, the link to resonances was made, the role of multiphoton transitions between field-dressed states and the resulting enhanced ionization was not discussed.
Our numerical results showed that the optical response is dominated by the ordinary Kerr effect and the ground-state depletion, i.e., the plasma-like response of Rydberg and free electrons. 
We were thus led to the conclusion that an augmented standard model may be obtained by replacing the Keldysh-like ionization rate in the plasma term by an ionization rate derived from non-Hermitian Floquet theory -- since in the adiabatic regime, Floquet theory accurately captures ionization enhancements due to multiphoton resonances.

The AFR model yielded excellent results in agreement with the ab initio TDSE calculation of the nonlinear refractive index. 
Our results also seem to rule out the necessity to introduce higher-order Kerr terms, since our modified standard model only contains the lowest order Kerr term $\propto n_2 I$. 
We therefore suggest that it is mainly resonance-enhanced ionization which led to the discrepancies between the experimental results of~\cite{Loriot2009} and the standard model predictions.
For a conclusive proof of our suggestion, though, it is essential to numerically reenact the experimental protocol of~\cite{Loriot2009}, which utilized a non-collinear pump-probe setup to detect the Kerr-induced birefringence. 
This requires full three-dimensional simulations of the TDSE and offers the chance for a detailed understanding of what was actually measured in the experiment. 

A further issue which has to be deepened in further research is the transition from adiabatic to non-adiabatic pulse envelopes, i.e., few-cycle pulses. 
As our calculations have shown, the ionization rate $\Gamma_0$ from the AFR calculations is not strictly valid in this regime due to smearing-out of internal resonances. 
While in the adiabatic regime the atomic population remains in the ground-state resonance, in the non-adiabatic case, transitions to higher-lying resonances have to be considered. 
Altogether, we expect that a promising path towards an efficient model of light-matter interaction consists of analyzing the interaction in a basis of complex-scaled Floquet resonances. 



\begin{acknowledgments}
  We thank Albert Ferrando, Felipe Morales and Maria Richter for valuable remarks on resonances in the nonlinear optical response.
  We also like to express our gratitude towards Matthias Wolfram and Oleh Omel'chenko for fruitful discussions about Floquet theory.
  Financial support by the  Deutsche Forschungsgemeinschaft, grant BR 4654/1-1, is gratefully  acknowledged.
\end{acknowledgments}


\bibliography{literature}

\begin{thebibliography}{10}

\bibitem{strickland}
D.~Strickland and G.~Mourou, \emph{Compression of amplified chirped optical
  pulses}, Optics Communications \textbf{56} 219  (1985).

\bibitem{A.1995}
A.~Braun, G.~Korn, X.~Liu, D.~Du, J.~Squier and G.~Mourou,
  \emph{{Self-channeling of high-peak-power femtosecond laser pulses in air}},
  Opt. Lett. \textbf{20} 73 (1995).

\bibitem{Gaeta2000}
A.~L. Gaeta, \emph{Catastrophic collapse of ultrashort pulses}, PRL \textbf{84}
  3582 (2000).

\bibitem{Husakou2001}
A.~V. Husakou and J.~Herrmann, \emph{{Supercontinuum Generation of Higher-Order
  Solitons by Fission in Photonic Crystal Fibers}}, Physical Review Letters
  \textbf{87} 203901 (2001).

\bibitem{Kolesik2002}
M.~Kolesik, J.~V. Moloney and M.~Mlejnek, \emph{{Unidirectional Optical Pulse
  Propagation Equation}}, Phys. Rev. Lett. \textbf{89} 283902 (2002).

\bibitem{L.2007}
L.~Berg{\'{e}}, S.~Skupin, R.~Nuter, J.~Kasparian and J.-P. Wolf,
  \emph{{Ultrashort filaments of light in weakly ionized, optically transparent
  media}}, Reports Prog. Phys. \textbf{70} 1633 (2007).

\bibitem{Keldysh1965}
L.~V. Keldysh, \emph{Ionization in the field of a strong electromagnetic wave},
  Sov. Phys. JETP \textbf{20} 1307 (1965).

\bibitem{Loriot2009}
V.~Loriot, E.~Hertz, O.~Faucher and B.~Lavorel, \emph{{Measurement of high
  order Kerr refractive index of major air components}}, Opt. Express
  \textbf{17} 13429 (2009).

\bibitem{Bejot2010}
P.~B{\'{e}}jot, J.~Kasparian, S.~Henin, V.~Loriot, T.~Vieillard, E.~Hertz,
  O.~Faucher, B.~Lavorel and J.-P. Wolf, \emph{{Higher-Order Kerr Terms Allow
  Ionization-Free Filamentation in Gases}}, Phys. Rev. Lett. \textbf{104}
  103903 (2010).

\bibitem{Ettoumi2010}
W.~Ettoumi, P.~B{\'{e}}jot, Y.~Petit, V.~Loriot, E.~Hertz, O.~Faucher,
  B.~Lavorel, J.~Kasparian and J.-P. Wolf, \emph{{Spectral dependence of
  purely-Kerr-driven filamentation in air and argon}}, Phys. Rev. A \textbf{82}
  033826 (2010).

\bibitem{Bree2012}
C.~Br{\'{e}}e, A.~Demircan, G.~Steinmeyer and C.~Bree, \emph{{Kramers-Kronig
  relations and high-order nonlinear susceptibilities}}, Phys. Rev. A
  \textbf{85} 33806 (2012).

\bibitem{Polynkin2011}
P.~Polynkin, M.~Kolesik, E.~M. Wright and J.~V. Moloney, \emph{{Experimental
  Tests of the New Paradigm for Laser Filamentation in Gases}}, Phys. Rev.
  Lett. \textbf{106} 153902 (2011).

\bibitem{Kosareva2011}
O.~Kosareva, J.-F. Daigle, N.~Panov, T.~Wang, S.~Hosseini, S.~Yuan, G.~Roy,
  V.~Makarov and S.~{Leang Chin}, \emph{{Arrest of self-focusing collapse in
  femtosecond air filaments: higher order Kerr or plasma defocusing?}}, Opt.
  Lett. \textbf{36} 1035 (2011).

\bibitem{Volkova2012}
E.~A. Volkova, A.~M. Popov and O.~V. Tikhonova, \emph{{Polarisation response of
  a gas medium in the field of a high-intensity ultrashort laser pulse: high
  order Kerr nonlinearities or plasma electron component?}} (2012).

\bibitem{Volkova2013}
E.~A. Volkova, A.~M. Popov and O.~V. Tikhonova, \emph{{Nonlinear polarization
  response of a gaseous medium in the regime of atom stabilization in a strong
  radiation field}}, J. Exp. Theor. Phys. \textbf{116} 372 (2013).

\bibitem{Richter2013}
M.~Richter, S.~Patchkovskii, F.~Morales, O.~Smirnova and M.~Ivanov, \emph{{The
  role of the Kramers-Henneberger atom in the higher-order Kerr effect}}, New
  J. Phys. \textbf{15} 083012 (2013).

\bibitem{Hofmann2015}
M.~Hofmann and C.~Br\'{e}e, \emph{{Femtosecond filamentation by intensity
  clamping at a Freeman resonance}}, Phys. Rev. A \textbf{92} 013813 (2015).

\bibitem{Kolesik2013}
M.~Kolesik and J.~V. Moloney, \emph{{Modeling and simulation techniques in
  extreme nonlinear optics of gaseous and condensed media.}}, Rep. Prog. Phys.
  \textbf{77} 016401 (2013).

\bibitem{Spott2014}
A.~Spott, A.~Jaron-Becker and A.~Becker, \emph{{Ab initio and perturbative
  calculations of the electric susceptibility of atomic hydrogen}}, Phys. Rev.
  A \textbf{90} 013426 (2014).

\bibitem{Lorin2012}
E.~Lorin, S.~Chelkowski, E.~Zaoui and A.~Bandrauk,
  \emph{{Maxwell-Schr{\"{o}}dinger-Plasma (MASP) model for laser-molecule
  interactions: Towards an understanding of filamentation with intense
  ultrashort pulses}}, Phys. D Nonlinear Phenom. \textbf{241} 1059 (2012).

\bibitem{Rensink2014}
T.~C. Rensink, T.~M. Antonsen, J.~P. Palastro and D.~F. Gordon, \emph{{Model
  for atomic dielectric response in strong, time-dependent laser fields}},
  Phys. Rev. A \textbf{89} 033418 (2014).

\bibitem{Kolesik2014}
M.~Kolesik, J.~M. Brown, A.~Teleki, P.~Jakobsen, J.~V. Moloney and E.~M.
  Wright, \emph{{Metastable electronic states and nonlinear response for
  high-intensity optical pulses}}, Optica \textbf{1} 323 (2014).

\bibitem{Freeman1987a}
R.~R. Freeman, P.~H. Bucksbaum, H.~Milchberg, S.~Darack, D.~Schumacher and
  M.~E. Geusic, \emph{{Above-threshold ionization with subpicosecond laser
  pulses}}, Phys. Rev. Lett. \textbf{59} 1092 (1987).

\bibitem{hhg}
C.~Zagoya, C.-M. Goletz, F.~Grossmann and J.-M. Rost, \emph{An analytical
  approach to high harmonic generation}, New Journal of Physics \textbf{14}
  093050 (2012).

\bibitem{Gavrila2008}
M.~Gavrila, I.~Simbotin and M.~Stroe, \emph{{Low-frequency atomic stabilization
  and dichotomy in superintense laser fields from the high-intensity
  high-frequency Floquet theory}}, Physical Review A - Atomic, Molecular, and
  Optical Physics \textbf{78} 1 (2008).

\bibitem{Su1996}
Q.~Su, B.~P. Irving, C.~W. Johnson and J.~H. Eberly, \emph{{Stabilization of a
  one-dimensional short-range model atom in intense laser fields}}, Journal of
  Physics B: Atomic, Molecular and Optical Physics \textbf{29} 5755 (1996).

\bibitem{Goldberg1967}
A.~Goldberg, \emph{{Computer-Generated Motion Pictures of One-Dimensional
  Quantum-Mechanical Transmission and Reflection Phenomena}}, Am. J. Phys.
  \textbf{35} 177 (1967).

\bibitem{NumRecipes1993}
W.~H. Press, S.~A. Teukolsky, W.~T. Vetterling and B.~P. Flannery,
  \emph{Numerical Recipes in FORTRAN -- The Art of Scientific Computing},
  Cambridge University Press, New York, 2nd edn. (1993).

\bibitem{Manolopoulos2002}
D.~E. Manolopoulos, \emph{{Derivation and reflection properties of a
  transmission-free absorbing potential}}, J. Chem. Phys. \textbf{117} 9552
  (2002).

\bibitem{Popov2003}
A.~M. Popov, O.~V. Tikhonova and E.~A. Volkova, \emph{{Strong-field atomic
  stabilization: numerical simulation and analytical modelling}}, J. Phys. B
  At. Mol. Opt. Phys. \textbf{36} R125 (2003).

\bibitem{Bejot2013}
P.~B\'{e}jot, E.~Cormier, E.~Hertz, B.~Lavorel, J.~Kasparian, J.-P. Wolf and
  O.~Faucher, \emph{{High-Field Quantum Calculation Reveals Time-Dependent
  Negative Kerr Contribution}}, Physical Review Letters \textbf{110} 043902
  (2013).

\bibitem{Fedorov1997}
M.~V. Fedorov, \emph{{Atomic and free electrons in a strong light field}},
  World Scientific, Singapore (1997).

\bibitem{Moiseyev1998}
N.~Moiseyev, \emph{{Quantum theory of resonances: calculating energies, widths
  and cross-sections by complex scaling}}, Physics Reports \textbf{302} 212
  (1998).

\bibitem{muller1999}
H.~G. Muller, \emph{{An Efficient Propagation Scheme for the Time-Dependent
  Schr{\"{o}}dinger Equation in the Velocity Gauge}}, Laser Phys. \textbf{9}
  138 (1999).

\bibitem{ABCtheorem1}
J.~Aguilar and J.~Combes, \emph{A class of analytic perturbations for one-body
  schr\"odinger hamiltonians}, Commun.~Math.~Phys. \textbf{22} 269 (1971).

\bibitem{ABCtheorem2}
E.~Balslev and J.~Combes, \emph{Spectral properties of many-body schr\"odinger
  operators with dilatation-analytic interactions}, Commun.~Math.~Phys.
  \textbf{22} 280 (1971).

\bibitem{Freeman1992}
G.~N. Gibson and R.~R. Freeman, \emph{Verification of the dominant role of
  resonant enhancement in short-pulse multiphoton ionization}, Phys.~Rev.~Lett.
  \textbf{69} 1904 (1992).

\bibitem{Brown2012}
J.~M. Brown, E.~M. Wright, J.~V. Moloney and M.~Kolesik, \emph{{On the relative
  roles of higher-order nonlinearity and ionization in ultrafast light-matter
  interactions}}, Opt.~Lett. \textbf{37} 1604 (2012).

\bibitem{replenishment}
M.~Mlejnek, E.~M. Wright and J.~V. Moloney, \emph{{Dynamic spatial
  replenishment of femtosecond pulses propagating in air}}, Optics Letters
  \textbf{23} 382 (1998).

\bibitem{Nurhuda2002}
M.~Nurhuda, A.~Suda and K.~Midorikawa, \emph{{Ionization-induced high-order
  nonlinear susceptibility}}, Phys. Rev. A \textbf{66} 1 (2002).

\bibitem{Kohler2013}
C.~K{\"{o}}hler, R.~Guichard, E.~Lorin, S.~Chelkowski, A.~D. Bandrauk,
  L.~Berg{\'{e}} and S.~Skupin, \emph{{Saturation of the nonlinear refractive
  index in atomic gases}}, Phys. Rev. A \textbf{87} 043811 (2013).

\end{thebibliography}

\end{document}